\documentclass[twocolumn,pra,superscriptaddress]{revtex4-2}
\usepackage{graphicx}
\usepackage{dcolumn}
\usepackage{bm}
\usepackage{mathtools}

\begin{document}

\title{Identifying quantum coherence in quantum annealers}

\author{Connor Aronoff}
\affiliation{Department of Physics and Astronomy, Texas Tech University, \\ Texas Tech University, Lubbock, TX 79409-1051, USA}
\email{coaronof@ttu.edu}
\author{Travis Howard}
\affiliation{Department of Physics and Astronomy, Texas Tech University, \\ Texas Tech University, Lubbock, TX 79409-1051, USA}
\author{David Nicholaeff}
\affiliation{New Mexico Consortium, Los Alamos, New Mexico 87544, USA} 
\author{Alejandro Lopez-Bezanilla}
\affiliation{Theoretical Division, Los Alamos National Laboratory, Los Alamos, New Mexico 87545, USA}
\author{Wade DeGottardi}
\affiliation{Department of Physics and Astronomy, Texas Tech University, \\ Texas Tech University, Lubbock, TX 79409-1051, USA}

\date{\today}

\begin{abstract}

Demonstrating genuine many-body quantum coherence in large-scale quantum processors remains a central challenge for near-term quantum technologies. Recent experiments on D-Wave quantum annealers have investigated quenches of Ising chains and observed defect densities that show Kibble–Zurek scaling, consistent with coherent quantum dynamics. However, identical scaling can arise from classical or thermal processes. Here we propose the use of many-body coherent oscillations (MBCO) as a  diagnostic for the identification of system-wide coherence in analog quantum simulators. Solving the time-dependent Schr\"odinger equation, we show that quenches of a staggered one-dimensional Ising chain across a quantum critical point produce oscillatory signatures in defect observables. We implement this model on the D-Wave Advantage quantum annealer. Using fast-anneal protocols, we find that, although defect densities follow Kibble–Zurek scaling, the expected oscillatory behavior is absent. We demonstrate that static disorder associated with individual qubits is not likely responsible for the absence of MBCO. Modest modifications to annealing schedules can dramatically enhance oscillation visibility. This work gives a general roadmap for the search for quantum coherence in noisy, large-scale quantum platforms.
\end{abstract}

\maketitle

The design of quantum information processors needs to reconcile the competing demands of coherence and scalability. While long coherence times can be achieved in small systems of well-isolated qubits, maintaining coherent dynamics in large, highly connected architectures remains a principal obstacle to scalable quantum computation~\cite{wendin_quantum_2017,gill_quantum_2022,de_leon_materials_2021}. Recent technological advances have enabled the fabrication of superconducting circuits containing thousands of controllable qubits, establishing a powerful platform for exploring collective quantum dynamics at scale. Among these architectures, quantum annealers occupy a distinctive niche: rather than implementing gate sequences, they exploit time-dependent Hamiltonian evolution and quantum tunneling to search for low-energy configurations of optimization problems~\cite{apolloni_quantum_1989,finnila_quantum_1994,kadowaki_quantum_1998,johnson_quantum_2011}. Achieving large connectivity and precise control in such systems, however, comes at the cost of increased exposure to decoherence arising from macroscopic circuitry, control lines, and flux noise~\cite{clarke_superconducting_2008,wendin_quantum_2017,bunyk_architectural_2014,chancellor_circuit_2017,das_decoherence_2009,nguyen_high-coherence_2019,tuokkola_methods_2025,ganjam_surpassing_2024}.

Assessing whether these devices sustain genuine many-body quantum coherence is therefore both essential and challenging. Recently, research on the D-Wave  Advantage quantum annealer, a machine with 4500 superconducting flux qubits, reported observations of coherent dynamics in a programmed one-dimensional transverse-field Ising chain~\cite{king_coherent_2022,king_computational_2024, bando_probing_2020}. These experiments involve quenching the system from a paramagnetic to a ferromagnetic phase, producing domain-wall defects whose density scales with anneal time in accordance with predictions of the Kibble–Zurek mechanism (KZM)~\cite{kibble_topology_1976,zurek_cosmological_1985,campo_universality_2014,huang_kibble-zurek_2014,chandran_kibble-zurek_2012}. While such scaling is consistent with quantum critical dynamics, the same power law can arise from classical processes. In fact, the diffusion–annihilation of domain walls in one dimension yields an identical power-law exponent~\cite{zwerger_critical_1981,krebs_finite-size_1995}, although a detailed statistical analysis presented in Ref.~\cite{king_coherent_2022} suggests that these processes alone cannot fully account for the results. Scaling behavior alone constitutes circumstantial rather than definitive evidence of quantum coherence. 

This ambiguity highlights a broader conceptual gap: what observables provide clear evidence of coherent many-body quantum dynamics in noisy, large-scale quantum simulators?
Oscillatory behavior, whether in space or time, is a hallmark of quantum coherence. Spatial Friedel oscillations reflect the coherence of fermionic quasiparticles near a Fermi surface~\cite{simion_friedel_2005}, while temporal St\"uckelberg oscillations arise when a system is driven through multiple quantum critical points, resulting in interference between distinct excitation pathways. The latter have been theoretically predicted in one-dimensional Ising chains~\cite{yan_nonadiabatic_2021, bayocboc_persistent_2025} where they manifest as oscillations in defect observables following a quench. The observation of these many-body coherent oscillations (MBCO) would provide a clear signature of quantumness.

In this work, we propose MBCO as a phase-sensitive diagnostic of quantum coherence in analog quantum annealers. We first articulate three criteria necessary for MBCO to exist and be observable on a quantum annealer. Focusing on a staggered one-dimensional Ising chain~\cite{igloi_quantum_1988,reichardt_quantum_2004,mishra_finite_2018,yan_nonadiabatic_2021}, we show theoretically that quenches across a quantum critical point generate oscillatory signatures in defect observables arising from many-body quantum interference, which are robust to static disorder. We implement this model on a quantum annealer and perform a systematic search for MBCO using fast-anneal protocols. While the measured defect densities reproduce Kibble–Zurek scaling, the absence of oscillatory signatures places quantitative bounds on coherent many-body dynamics under current operating conditions and highlights the need for interference-based analysis beyond scaling laws to certify quantumness in large-scale quantum devices.

\begin{figure}[t!]
\centering
\includegraphics[scale=0.52]{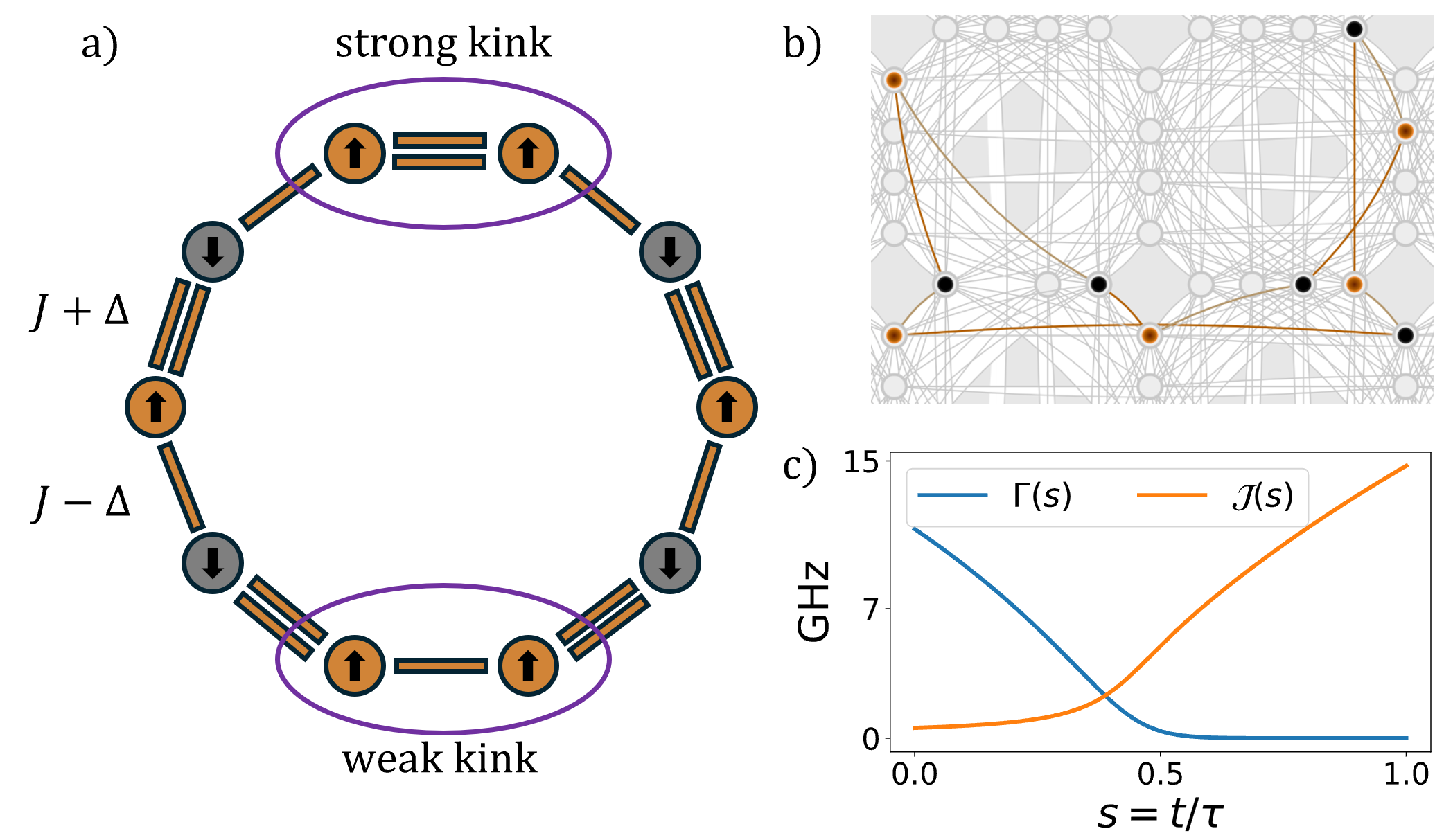}
\caption{(a) The staggered Ising chain with alternating strong ($J + \Delta$) and weak ($J -\Delta$) bond strengths. For an antiferromagnetic chain, kinks correspond to neighboring spins that are parallel after a quench. (b) Embedding of a staggered 10-qubit Ising chain on the D-Wave Advantage machine. The coloring of the qubits indicate the results of an anneal, with spin-up and spin-down indicated by orange and gray, respectively. (c) D-Wave annealing schedule~\cite{noauthor_annealing_nodate} for the couplings $\mathcal{J}(s)$ and $\Gamma(s)$ in Hamiltonian (\ref{eq:hamiltonian}) as a function of the anneal parameter $s = t/\tau$.}
\label{fig:staggeredchain}
\end{figure}

Before describing our search for MBCO, we identify several key prerequisites for their detection. (1) The annealing time $\tau$ must not exceed the qubit coherence time, which is $5$–$50$ nsec for D-Wave~\cite{noauthor_annealing_nodate}. (2) The detection of oscillations of frequency $\Omega$ requires data with a sampling precision $\Delta\tau$ that satisfies the Nyquist–Shannon condition $\Omega < 1/2\Delta\tau$ \cite{nyquist_certain_1928, shannon_communication_1949}. (3) The energy of the MBCO must exceed $k_B T$. A temperature $T^\ast \lesssim \hbar \Omega / k_B$, where $k_B$ and $h$ are the Boltzmann and Planck constants, allows for the detection of a single quanta in principle. Conditions (1)-(3) are satisfied using D-Wave's fast-anneal protocol. Condition (1) is satisfied given anneals as short as $\tau \approx 5~\mathrm{nsec}$ are accessible. The MBCO considered here have a frequency $\Omega \approx 5$ GHz. For annealing times $\tau \sim 10$ nsec, the precision $\Delta \tau \sim 20$ psec, which, according to the Nyquist-Shannon condition (2), allows the detection of oscillations $\Omega \lesssim 25$ GHz. Condition (3) gives a maximum temperature of $T^\ast\approx 250$ mK for $\Omega \approx 5$ GHz, while D-Wave operates at $12$-$15$ mK. 

\begin{figure*}[t!]
\centering
  \includegraphics[scale=0.5]{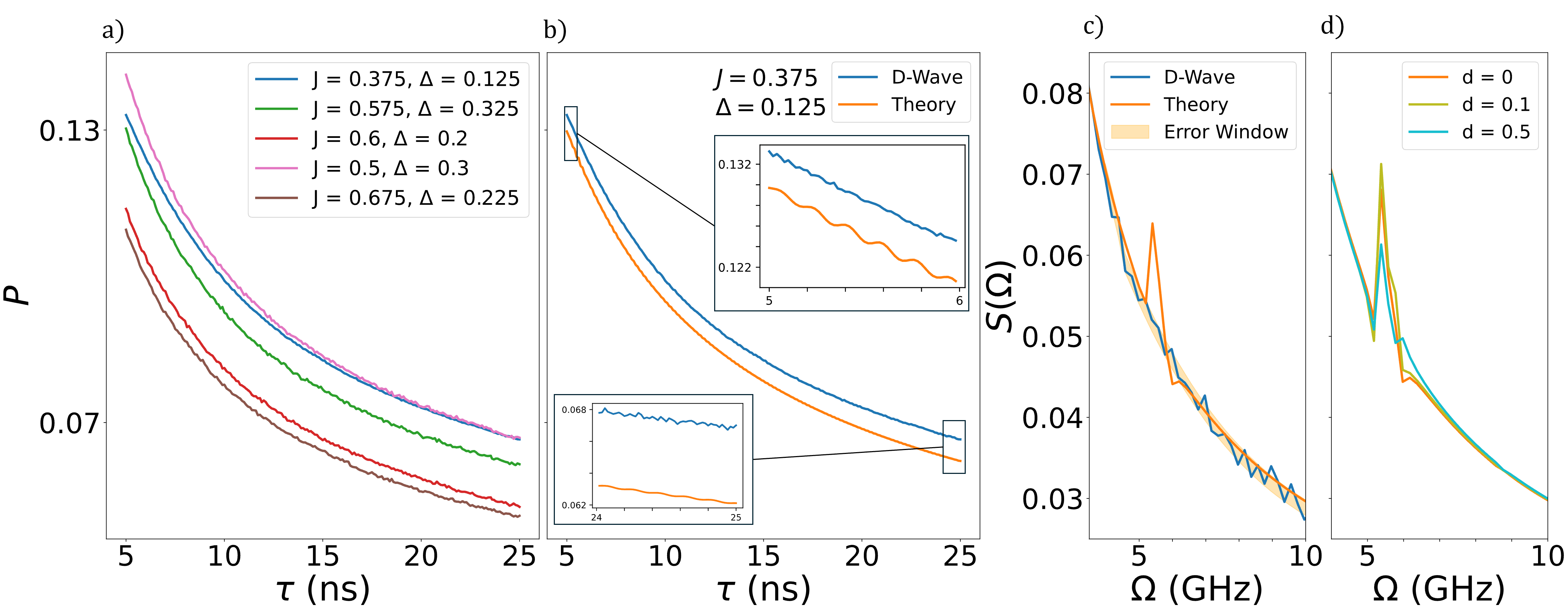}
\caption{(a) The average difference in kink densities across weak and strong bonds $P(\tau)$ for quenches of staggered Ising chains of length $L=160$ on using D-Wave anneals. Each line represents the mean of approximately 75,000 anneals for the indicated bond strengths $J$ and $\Delta$. 
(b) $P(\tau)$ for D-Wave (blue) and theory (orange). Magnified insets show theoretically predicted many-body coherent oscillations are most pronounced early in the anneal. These oscillations are absent in the D-Wave data. (c) Fourier transform $S(\Omega)$ of $P(\tau)$ for D-Wave data and theory from (b). Theoretical curve for $S(\Omega)$ exhibits a well-defined peak at $\Omega \approx 5.4~\mathrm{GHz}$. The error window (orange band) has a full width corresponding to 2$\sigma$.
(d) Fourier transform $S(\Omega)$ for the theoretical model accounting for disorder in Zeeman field, where $d$ is the strength of the disorder. We find that the prominent feature at $\Omega \approx 5.4~\mathrm{GHz}$ is robust to strong disorder.}
\label{fig:theory_dwave_comp}
\end{figure*}

The time-dependent Hamiltonian of the staggered Ising chain~\cite{reichardt_quantum_2004, mishra_finite_2018} is
\begin{equation}
H(s)/h  = \mathcal{J}(s)\sum_{n=0}^N(J+\Delta(-1)^n)\sigma_{n}^{z}\sigma_{n+1}^{z} + \Gamma(s)\sum_{n}\sigma_{n}^{x},
\label{eq:hamiltonian}
\end{equation}
where the annealing parameter $s = t/\tau$, $t$ is time, $\tau$ is the duration of the anneal, and $\sigma_n^x$ and $\sigma_n^z$ denote the Pauli spin operators for the $n^{th}$ qubit. The time-dependent frequencies $\mathcal{J}\left(s\right) \geq 0$ and $\Gamma(s) \geq 0$ set the relative strength of the ferromagnetic and paramagnetic terms, while the dimensionless parameters $J, \Delta > 0$ set the staggering strength. As shown in  Figure~\ref{fig:staggeredchain}a, the Hamiltonian (\ref{eq:hamiltonian}) is characterized by alternating strong and weak couplings proportional to $J+\Delta$ and $J-\Delta$, respectively.

Figure~\ref{fig:staggeredchain}b shows the D-Wave annealing schedules for the transverse field $\Gamma(s)$ and the Ising coupling $\mathcal{J}(s)$. At the start of the anneal, $\mathcal{J}(0)=0$ and $\Gamma(0) \approx 11$ GHz, deep in the paramagnetic phase. At the end of the anneal the system is in the ferromagnetic phase with $\mathcal{J}(1) \approx 15$ GHz and $\Gamma(1)=0$. The boundary between these two phases is given by
\begin{equation}
\sqrt{J^2 + \Delta^2} \mathcal{J}(s) = \Gamma(s)
\label{eq:critical}
\end{equation}
\cite{zhang_exact_1994, yan_nonadiabatic_2021}.
Near the critical point, quantum fluctuations give rise to topological defects, i.e., domain-wall kinks. For the antiferromagnetic chain, these defects are associated with coupled spins that are parallel after the anneal, see Figure~\ref{fig:staggeredchain}a. The expectation value of the total defect density, which is the number of defects per qubit, is 
$K(\tau) =\frac{1}{2N} \sum_{n} \left( 1 + \langle \sigma_{n}^{z}\sigma_{n+1}^{z} \rangle_{s=1} \right)$ at the end of the anneal $(s = 1)$. The difference in the kink densities across weak and strong bonds is
\begin{equation}
P(\tau) = \frac{1}{2 N} \sum_{n} (-1)^n \langle \sigma_{n}^{z}\sigma_{n+1}^{z} \rangle_{s=1}.
\end{equation}
The quantity $P > 0$ since fewer defects are expected across strong bonds ($J + \Delta$) than weak bonds ($J-\Delta$). 

The staggered Ising chain can be solved exactly via a Jordan-Wigner transformation in which spinless fermions $\psi_{n} = \frac{1}{2}\prod_{j=1}^{n-1}\sigma_{j}^{x}(\sigma_{n}^{z} + i\sigma_{n}^{y})$ are introduced at each site $n$~\cite{igloi_quantum_1988, zhang_exact_1994, yan_nonadiabatic_2021}. The resultant fermionic Hamiltonian is quadratic (see the Supplementary Material). In $k$-space, the Hamiltonian is block-diagonal. For the initial conditions considered here, these blocks can be taken to be $6 \times 6$ matrices. Calculating the number of defects is thus reduced to solving the time-dependent Schr\"{o}dinger equation for each of these blocks. For sufficiently long quenches, the total number of kinks follows Kibble-Zurek scaling~\cite{huang_kibble-zurek_2014}. The theoretical prediction for the difference in kink densities $P(\tau)$ across strong and weak bonds for particular values of $J$ and $\Delta$ is shown in Figure~\ref{fig:theory_dwave_comp}b for short annealing times. The number of defects is found to oscillate, with some beating, at a frequency $\Omega \sim \mathcal{J}(s_c) \Delta$, where $s_c \approx 1/3$ is the annealing parameter at criticality. The Fourier transform of $P(\tau)$, shown in Figure~\ref{fig:theory_dwave_comp}c, reveals a sharp feature at $\Omega \approx 5.4$ GHz. The sensitivity of $S(\Omega)$ to the range of annealing times is discussed in the Supplementary Material.

We conducted a systematic search for MBCO on the D-Wave Advantage system. Our quantum simulations implemented fast quenches of multiple realizations of the staggered Ising chain embedded as rings of 160 qubits, as shown in Figure~\ref{fig:embedding}. Before acquiring data on D-Wave, we carried out a calibration procedure known as shimming to account for variations associated with individual qubit biases and coupler strengths. For details, refer to Figure~\ref{fig:shim} of Supplementary Material. In total, 90,000 fast anneals processes were conducted for each annealing time $\tau$. The first 15,000 anneals were used for the shimming process and 75,000 anneals contributed to the usable data. 

The curves in Figure~\ref{fig:theory_dwave_comp}a show $P(\tau)$ obtained from these $75{,}000$ anneals for 200 different annealing times $\tau$. A discrete Fourier transform of the data shown in Figure~\ref{fig:theory_dwave_comp}b yields the jagged blue curve in Figure~\ref{fig:theory_dwave_comp}c. To quantify the statistical uncertainty in the Fourier amplitudes $S(\Omega)$, the data was partitioned into $75{,}000$ artificial time series by random assignment. Each series was Fourier transformed, and the standard deviation of the spectra defined the frequency-resolved noise level $\sigma(\Omega)$. The shaded band in Figure~\ref{fig:theory_dwave_comp}c indicates the $2\sigma$ window about the best-fit noise baseline. Repeating the randomization procedure with different time series produced indistinguishable noise estimates, confirming the robustness of the extracted $\sigma(\Omega)$.

Despite the fact that our search was conducted in D-Wave's putative coherent regime, the collected data shows no evidence of the predicted oscillations, although the measured defect densities are consistent with the predictions of Kibble–Zurek (KZ) theory (see Supplementary Material).
The spectrum $S(\Omega)$, obtained by Fourier transforming the analytical-derived curve in Figure~\ref{fig:theory_dwave_comp}b, exhibits a peak near $\Omega \approx 5.4~\mathrm{GHz}$ and is more than $5\sigma$ above the center of the noise window (yellow band).
The Fourier transform of the D-Wave data shows no discernible structure at this frequency. This allows us to state with high confidence that the predicted oscillations are absent in the annealer-derived data.

\begin{figure*}[t!]
\centering\includegraphics[scale=0.5]{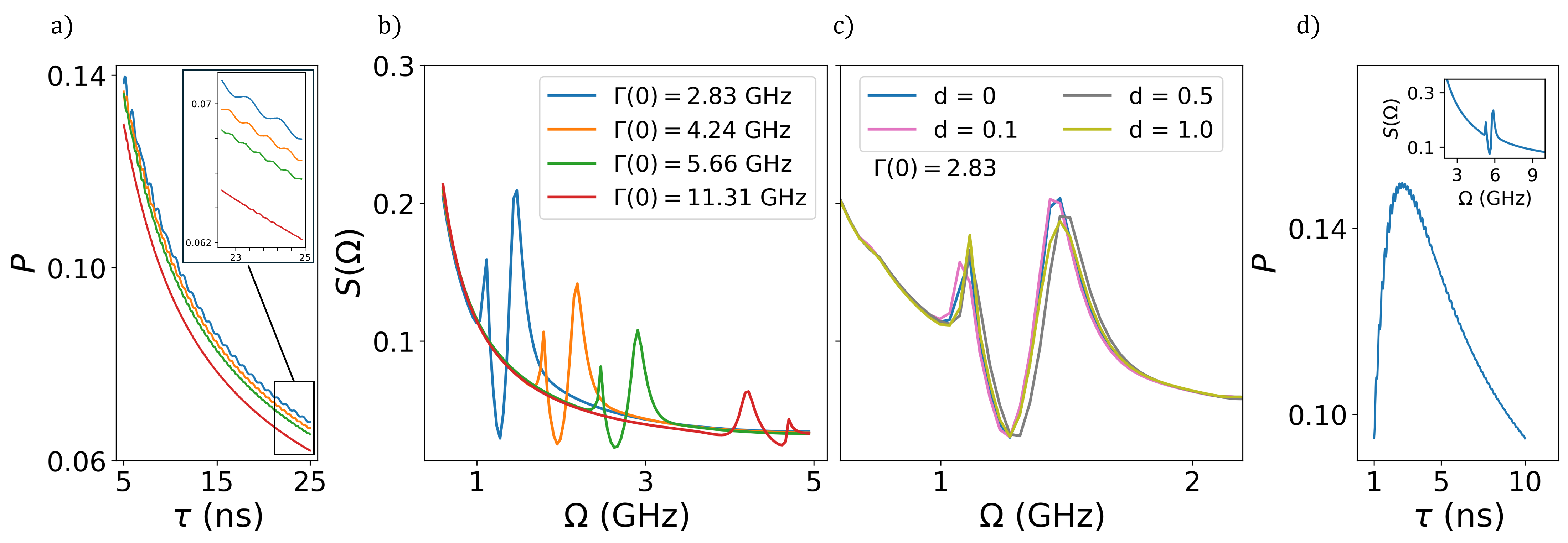}
\caption{Theoretical results for annealing parameters beyond the range currently accessible on D-Wave. (a) Calculated expectation value of the difference in kink densities $P$ as a function of annealing time $\tau$ for various initial strengths of the parameter $\Gamma(0)$ (see legend in (b)). For these values of $\Gamma(0)$, a doublet peak structure in $S(\Omega)$ is visible. (b) Peaks in the Fourier transform of $S(\Omega)$ of $P$ become more prominent as $\Gamma(0)$ is decreased, in agreement with (a). (c) $S(\Omega)$ for $\Gamma(0)=2.83$~GHz and various disorder strengths $d$. (d) $P(\tau)$ for ultra-short anneal times ($\tau<5$~ns). (Inset) Fourier transform $S(\Omega)$ of $P(\tau)$.}
\label{fig:wishlist_fig}
\end{figure*}

A potential source of the discrepancy between theory and experiment is the static disorder in the quantum annealer, which could suppress coherent oscillations. Such disorder can arise from variations in the flux qubits and their couplings to control lines. Disorder of this type can be partially mitigated through shimming, a calibration procedure that reduces the spread in the effective programmed couplings $\mathcal{J}(s)J_n$ and anneal offsets (see Supplementary Material), where $J_n$ is the effective coupling strength for the $n^{th}$ qubit. However, there is no way to address any potential disorder in the Zeeman-like terms $\sigma_n^x$. To investigate its impact, we consider a generalization of our Hamiltonian in which Zeeman-like terms in Eq.~(\ref{eq:hamiltonian}) are replaced by $\Gamma(s)\sum_n h_n \sigma_n^x$, with $h_n = 1 + dx_n$ and $x_n$ is drawn uniformly from $[-1,1]$. The disorder gives rise to terms in the Hamiltonian of the form $h_{2k}\,\psi_k^\dagger \psi_{-k}^{\phantom\dagger}$ in the corresponding fermionic theory, where $\psi_k$ is the annihilation operator of a fermion with wave number $k$ and $h_{2k}$ is the Fourier transform of $h_n$. These terms describe backscattering of the fermionic quasiparticles of the system. Remarkably, as shown in Figure~\ref{fig:theory_dwave_comp}, the many-body coherent oscillations remain robust over a wide range of disorder strengths $d$, and persist even as the disorder $d$ approaches $0.5$. This result, together with our detailed shimming procedure, rules out static disorder as a likely cause for the absence of coherent oscillations in the D-Wave data. 

Although our theory predicts that the MBCO should in principle be observable on the D-Wave platform, their visibility is small (see Figure~\ref{fig:theory_dwave_comp}b). We therefore propose modifications of the annealing schedule that substantially enhance the visibility of the MBCO. One way to accomplish this is to reduce the scale of the transverse field $\Gamma(s)$ from its current value $\Gamma(0)\approx 11~\mathrm{GHz}$. Theoretical curves for $P(\tau)$ for several values of $\Gamma(0)$ are shown in Figure~\ref{fig:wishlist_fig}a. A second modification involves shortening the annealing time $\tau$. As shown in Figure~\ref{fig:wishlist_fig}b, reducing it to $\tau \approx 2.5~\mathrm{ns}$ leads to increased visibilities. St\"uckelberg oscillations are typically associated with quenches in which several quantum critical points are crossed. A likely reason for the small oscillations here is that only one quantum critical point is crossed during the anneal. In so-called ``reverse anneals,'' it may be possible to cross more than one critical point, however fast reverse anneals are not currently available to D-Wave users.

We have shown that a staggered one-dimensional Ising chain provides an analytically controlled and experimentally accessible diagnostic for probing many-body quantum coherence in large-scale quantum simulators. The quench dynamics of this system generate many-body coherent oscillations arising from quantum interference and offer a sophisticated test for many-body quantum coherence. While the rigorous requirements for observing these oscillations---fast control, high temporal resolution, and sustained phase coherence---are, in principle, met by recently introduced fast-anneal capabilities, our searches reveal no evidence of the predicted oscillatory signatures on the staggered Ising chain. Furthermore, our analysis rules out static disorder as a likely culprit. These findings underscore a central challenge for quantum annealing architectures: achieving precise, system-wide synchronization across thousands of qubits during time-dependent evolution. For example, even small mismatches in timing or control amplitudes can induce collective dephasing, suppressing the many-body interference effects that distinguish genuine quantum dynamics from classical or incoherent behavior.

Beyond diagnosing current limitations, our results establish a clear roadmap for future progress. We demonstrate that modest modifications to annealing schedules can dramatically enhance the visibility of many-body coherent oscillations, suggesting concrete directions for hardware and control improvements. Specifically, our results point to greater schedule flexibility and reduced transverse-field scales as a way of accessing coherent many-body interference in quantum annealers. More broadly, this work highlights that power-law defect scaling alone is insufficient to certify quantumness in noisy, large-scale quantum systems. Instead, interference-based observables such as MBCO provide a sharp signature of many-body coherence. Targeted searches for these oscillations therefore offer a powerful and broadly applicable framework for validating quantum coherence in next-generation analog quantum simulators.

\vspace{0.3 in}

Los Alamos National Laboratory is managed by Triad National Security, LLC, for the National Nuclear Security Administration of the U.S. Department of Energy under Contract No. 89233218CNA000001. The authors thank the New Mexico Consortium, under subcontract C2778, the Quantum Cloud Access Project (QCAP), for providing quantum computing resources and technical collaboration. We thank  the Institute for Materials Science (IMS) at Los Alamos for financial support; work was also supported by under the program of “IMS Rapid Response.” We gratefully acknowledge discussions with Gabe Schumm, Anders W. Sandvik, and Chris Rich. We are especially grateful to Andrew King for numerous enlightening discussions.

\bibliography{stagger_bib}
\bibliographystyle{naturemag}

\cleardoublepage
\onecolumngrid

\section{Supplemental Material}

\subsection{Exact Solution for the Staggered Ising Chain}

The Hamiltonian of the staggered Ising chain is
\begin{equation}
H = \sum_{n}(J+\Delta(-1)^n)\sigma_{n}^{z}\sigma_{n+1}^{z} + h\sum_{n}\sigma_{n}^{x}.
\label{eq:initial_ham}
\end{equation}
We apply a Jordan-Wigner transformation
\begin{equation}
    \psi_{n} = \frac{1}{2}\prod_{j=1}^{n-1}\sigma_{j}^{x}(\sigma_{n}^{z} + i\sigma_{n}^{y}). \quad
\end{equation}
The fermion operators obey anti-commutation relations $\{\psi^\dagger_{n}, \psi_{n'}\} = \delta_{nn'}$. In terms of the fermions, the Hamiltonian~(\ref{eq:initial_ham}) is
\begin{equation}
    H = \sum_{n}\left[J+\Delta(-1)^n \right](\psi_{n}\psi_{n+1} - \psi_{n}^{\dagger}\psi_{n+1}^{\dagger} - \psi_{n}^{\dagger}\psi_{n+1}^{\phantom\dagger} - \psi_{n+1}^{\dagger}\psi_{n}^{\phantom\dagger}) + h(2\psi_{n}^{\dagger}\psi_{n} - 1).
\end{equation}
Going to $k$-space, we have $\psi_{n} = \frac{1}{\sqrt{N}}\sum_{k}e^{ikn}\psi_{k}$, which gives
\begin{gather}
    H = \sum_{k} J(e^{-ik}\psi_{k}\psi_{-k} + e^{-ik}\psi_{k}^{\dagger}\psi_{-k}^{\dagger} 
    - e^{ik}\psi_{k}^{\dagger}\psi_{k}^{\phantom\dagger} -e^{-ik}\psi_{k}^{\dagger}\psi_{k}^{\phantom\dagger}) \notag \\
    + \Delta(-e^{-ik}\psi_{k}\psi_{\pi-k} + e^{-ik}\psi_{k}^{\dagger}\psi_{\pi-k}^{\dagger} 
+ e^{ik}\psi_{k}^{\dagger}\psi_{\pi+k}^{\phantom\dagger} - e^{-
    ik}\psi_{k}^{\dagger}\psi_{\pi+k}^{\phantom\dagger}) \notag \\
+ h(2\psi_{k}^{\dagger}\psi_{k}^{\phantom\dagger} - 1). \notag
\label{eq:hamiltonain_opp}
\end{gather}
The allowed values of $k = 2 \pi j/N$, for $0 \leq j \leq N$. The Hamiltonian couples fermion modes with wave numbers $k$, $-k$, $\pi + k$, and $\pi - k$. We work in the basis  $|n_k, n_{-k}, n_{\pi-k}, n_{\pi+k}\rangle$ where $n_k = \psi^\dagger_k\psi^{\phantom\dagger}_k$. The initial state of the system is $| 0, 0, 0, 0 \rangle$. For this initial state, only six states are accessible: $|1,1,1,1\rangle$, $|1,1,0,0\rangle$, $|1,0,1,0\rangle$, $|0,1,0,1\rangle$, $|0,0,1,1\rangle$ and $|0,0,0,0\rangle$.

Thus, the Hamiltonian can be expressed in terms of $6 \times 6$ blocks,
\begin{equation}
H = \sum_{0 < k < \pi/2} \hat{\Psi}_{k}^{\dagger}H_k^{\phantom\dagger}\hat{\Psi}_{k}^{\phantom\dagger},
\label{eq:full_ham}
\end{equation}
where now the allowed values of $k$ are restricted to the range indicated and
\begin{equation*}
    H_k =
       \begin{bmatrix}
			4h              & -2Ji\sin{k}       & -2\Delta \cos{k} & -2\Delta \cos{k} & -2Ji\sin{k}           & 0 \\
			2Ji\sin{k}      &  -4J\cos{k}       & -2i\Delta \sin{k} & -2i\Delta \sin{k}  & 0               & -2iJ\sin{k} \\
			-2\Delta \cos{k} & 2i\Delta \sin{k}  & 0               & 0                   & -2i\Delta \sin{k}  & 2\Delta \cos{k} \\
			-2\Delta \cos{k} & 2i\Delta \sin{k} & 0               & 0                   & -2i\Delta \sin{k}  & 2\Delta \cos{k} \\
			2Ji\sin{k}     & 0                 & 2i\Delta \sin{k} & 2i\Delta \sin{k}  & 4J\cos{k}          & -2iJ\sin{k} \\
			0              & 2iJ\sin{k}        & 2\Delta \cos{k} & 2\Delta \cos{k}   & 2iJ\sin{k}         & -4h
     \end{bmatrix},
\end{equation*}
and
\begin{equation}
\Psi^\dagger = \left( \langle1,1,1,1|, \langle 1,1,0,0 |, \langle 1,0,1,0 |, \langle 0,1,0,1|, \langle 0,0,1,1|, \langle0,0,0,0| \right).
\end{equation}

\begin{figure}[h!]
\centering
\includegraphics[scale=0.3]{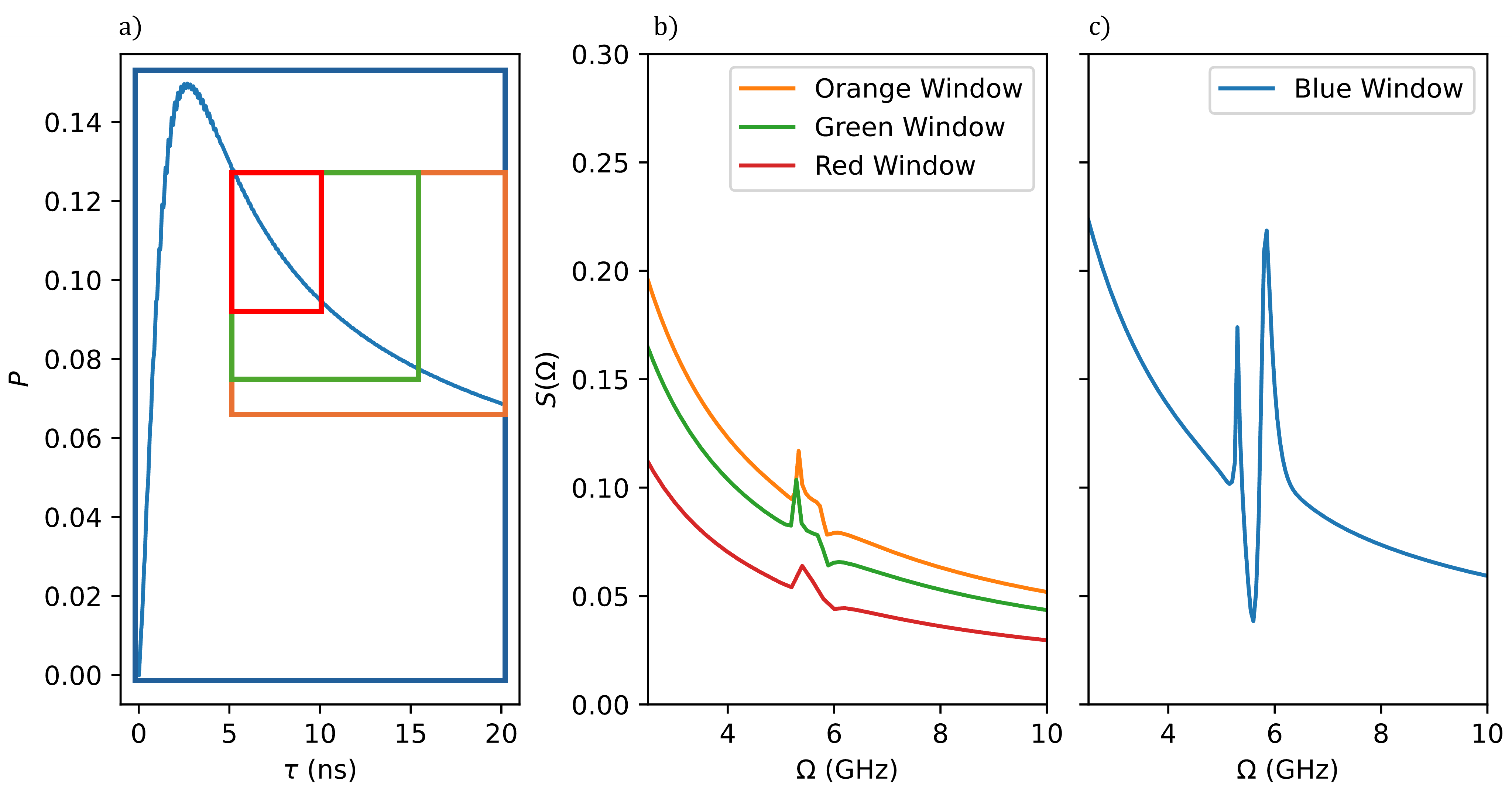}
\caption{Sensitivity of $S(\Omega)$ to the range of annealing times included in the Fourier transform. (a) Windows indicate range of annealing times used to calculate the corresponding $S(\Omega)$ curves in (b). The curves in Fig.~\ref{fig:theory_dwave_comp}c of the main text were obtained using the red window. (c) The double peak structure is most clearly visible for the window $0 \leq \tau \leq 20$ ns.}
\label{fig:fine}
\end{figure}

Theoretical curves for the Fourier transform of $P(\tau)$ are shown in Fig.~\ref{fig:fine} for various ranges of annealing times. This range controls the frequency resolution $\Delta \Omega$ of the $S(\Omega)$ curves.

\subsection{Disorder in the Transverse Field Calculation}

To account for disorder in the Zeeman-like terms, we consider introduce a term $H_\Gamma=\Gamma(s)\sum_{n}h_n \sigma_n^x$ in the Hamiltonian. In terms of fermions, this term becomes 
\begin{equation}
    H_{\Gamma} = \Gamma(s)\sum_{n}h_n(2\psi_{n}^{\dagger}\psi_{n}^{\phantom\dagger} - 1).
\end{equation}
In $k$-space, we have
\begin{equation}
H_\Gamma = \frac{\Gamma(s)}{\sqrt{N}}\sum_{k,q} h_{q}\psi^{\dagger}_{k} \psi_{k-q}^{\phantom\dagger},
\end{equation}
where $h_n = \frac{1}{\sqrt{N}}\sum_k e^{ikn} h_{k}$.
The terms with $q \approx 0$ and $q \approx 2k$ describe elastic scattering and thus dominate for annealing times $\tau$ that greatly exceed the characteristic period of the fast oscillations of the system $1/\Omega$. In other words, we ignore the rapidly oscillating terms in the Hamiltonian. What remains is
\begin{equation}
H_\Gamma = \frac{\Gamma(s)}{\sqrt{N}}\sum_{0<k<\pi/2}h_{2k}(\psi^{\dagger}_{k}\psi_{-k}^{\phantom\dagger} + \psi^{\dagger}_{-k}\psi_{k}^{\phantom\dagger} + \psi^{\dagger}_{\pi-k}\psi_{\pi+k}^{\phantom\dagger} + \psi^{\dagger}_{\pi+k}\psi_{\pi-k}^{\phantom\dagger}).
\end{equation}
These terms now allow the system to access the states $|1,0,0,1\rangle$ and $|0,1,1,0\rangle$ as well. The calculations of $P(\tau)$ for the disordered case thus involve diagonalizing each of the $8 \times 8$ blocks.

\subsection{Calibration and shimming for the staggered Ising chain}

We follow the iterative shimming methodology of Chern \emph{et al.}~\cite{chern_tutorial_2023}, adapting it to the symmetry of our one–dimensional staggered Ising chain. The goal is to compensate residual qubit biases and coupler miscalibrations so that, for a given programmed Hamiltonian, the sampled spin configurations respect the expected symmetries of the model.

\paragraph*{Symmetry orbits and target observables.}
The embedded chain has a discrete dihedral symmetry with two natural qubit orbits (even and odd sublattices) and two coupler orbits (strong bonds with target coupling $J_{\mathrm s}=J+\Delta J$ and weak bonds with $J_{\mathrm w}=J-\Delta J$). In zero longitudinal field ($h_i=0$) the Ising Hamiltonian is invariant under a global spin flip, so the ensemble-averaged single–qubit magnetization must vanish on each qubit orbit, $\langle \sigma_i^z\rangle_{\mathcal O_q}=0$. For antiferromagnetic couplers we target antialigned spin pairs, i.e., large negative correlations $\langle \sigma_i^z\sigma_j^z\rangle$, with statistics that are uniform within each coupler orbit. We quantify deviations using: (i) the orbit–averaged magnetizations $m_{\mathcal O_q}=\langle \sigma_i^z\rangle_{\mathcal O_q}$ and (ii) the \emph{frustration probability} for an antiferromagnetic coupler $(i,j)$,
\begin{equation}
p^{\mathrm{frust}}_{ij}=\frac{1+\operatorname{sgn}(J_{ij})\,\langle \sigma_i^z\sigma_j^z\rangle}{2},
\end{equation}
together with its dispersion within each orbit, $\mathrm{Std}_{\mathcal O_c}[p^{\mathrm{frust}}]$.

\paragraph*{Gauge averaging and iteration loop.}
Each iteration consists of: (1) programming a diagnostic instance, (2) collecting $\sim 10^4$ samples over multiple spin–reversal gauges to suppress systematic offsets, (3) computing the orbit–wise observables $\{m_{\mathcal O_q},\,p^{\mathrm{frust}}_{ij}\}$, and (4) updating local fields and couplers using small corrective ``shims.'' We employ simple proportional (gradient–descent–like) updates that respect the orbits.

\emph{Qubit (flux–bias) shim.}
We compensate residual single–qubit offsets using the QPU's per–qubit flux–bias offset controls. Concretely, we apply small corrections $\{\phi_i\}$ through the sampler’s \texttt{flux\_biases} parameter and update them orbitwise to drive the ensemble magnetizations toward their symmetry targets:
\begin{equation}
\phi_i \leftarrow \phi_i - \eta_\phi\, m_{\mathcal O_q}, \qquad i\in \mathcal O_q,
\end{equation}
with a learning rate $\eta_\phi>0$ and gauge averaging at each iteration. $\mathcal{O}_q$ denotes a qubit orbit, i.e., an equivalence class of qubits related
by automorphisms of the embedded Ising graph that preserve the Hamiltonian.
Using flux–bias offsets directly targets persistent device–level biases and drift while preserving the programmed problem Hamiltonian; it also aligns with the recommended shimming workflow for Advantage–generation processors~\cite{chern_tutorial_2023,noauthor_virtualgraphcomposite_nodate,noauthor_errors_nodate}. 

\emph{Coupler (frustration) shim.} To homogenize the effective couplings within each orbit and preserve the intended strong/weak ratio, we apply two complementary adjustments:
\begin{align}
J_{ij} &\leftarrow J_{ij}\!-\eta_J\, \delta_{\mathcal O_c}(i,j), \quad (i,j)\in\mathcal O_c,\\
\delta_{\mathcal O_c}(i,j) &=
p^{\mathrm{frust}}_{ij}-\overline{p^{\mathrm{frust}}}_{\mathcal O_c}
\end{align}
which suppress the intra–orbit dispersion of frustration probabilities. A coupler orbit $\mathcal{O}_c$ consists of all couplers that are symmetry-equivalent under the same automorphism group.

These two steps equalize couplers within orbits and enforce the intended dimerization across orbits. The learning rates $\eta_J$ are kept small at each annealing time to ensure stable convergence.
 
\begin{figure}[h!]
\centering
\includegraphics[scale=0.5]{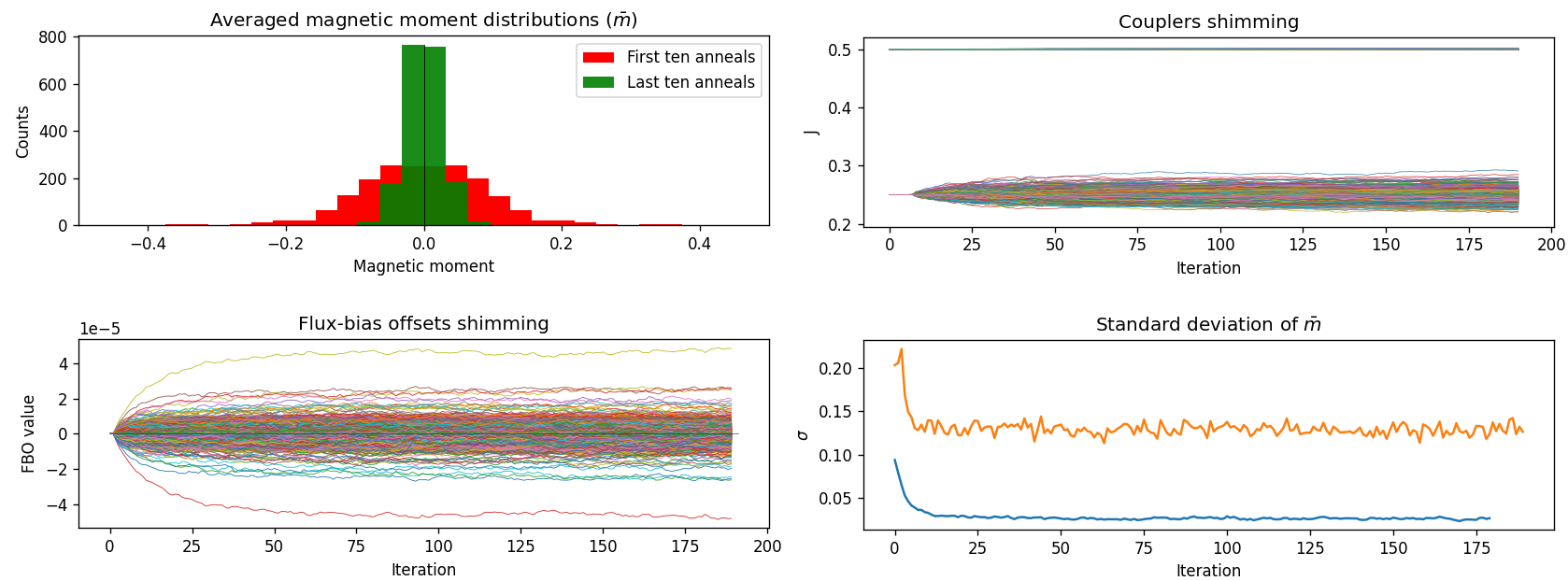}
\caption{Results of the iterative shimming procedure applied to the staggered Ising chain for the particular annealing time of $19 ns$. a) Histograms of the averaged magnetization $\tilde{m}$ for the first ten (red) and last ten (green). The distribution narrows and centers around zero as flux-bias offsets and coupler strengths are refined. b) Evolution of the programmed coupler strengths $J_{ij}$ over successive iterations converging toward stable values.
c) Convergence of per-qubit flux-bias offset (FBO) values for gradual suppression of residual local-field biases. d) Standard deviation $\sigma_{\tilde{m}}$ of the orbit-averaged magnetization as a function of iteration, indicating rapid stabilization of the ensemble.
Together, these panels illustrate how successive flux-bias and coupler adjustments restore symmetry between qubit orbits and minimize frustration, yielding a calibrated embedding faithful to the intended staggered Ising Hamiltonian.}
\label{fig:shim}
\end{figure}

\paragraph*{Convergence diagnostics.}
Figure~\ref{fig:shim}(a) shows histograms of the total magnetization over successive iterations: the initially broad, asymmetric distribution collapses toward a narrow, centered peak after shimming. Figure~\ref{fig:shim}(c) displays the evolution of $p^{\mathrm{frust}}_{ij}$ across all couplers; the spread within each orbit shrinks iteration by iteration. The corresponding orbit–wise standard deviation $\mathrm{Std}_{\mathcal O_c}[p^{\mathrm{frust}}]$, plotted in Figure~\ref{fig:shim}(d), serves as a scalar convergence metric. Once both $|m_{\mathcal O_q}|$ and $\mathrm{Std}_{\mathcal O_c}[p^{\mathrm{frust}}]$ fall below preset thresholds, qubit magnetizations and pair correlations stabilize and the shim is frozen.

\paragraph*{Remarks on the staggered chain.}
Specializing to the dimerized chain helps in two ways. First, the symmetry reduction to two qubit orbits and two coupler orbits yields high measurement statistics per orbit, accelerating convergence. Second, the explicit weak/strong structure provides an internal consistency check, namely after shimming, the hardware–measured ratio of orbit–averaged correlations (or linear response slopes) must match the programmed dimerization to within statistical error. We find that combining qubit–bias and coupler–frustration shims yields a dual calibration of local fields and interactions that is necessary to prevent spurious lifting of degeneracies or distortion of defect sampling in subsequent quench experiments.

\begin{figure}[h!]
\centering
\includegraphics[scale=0.2]{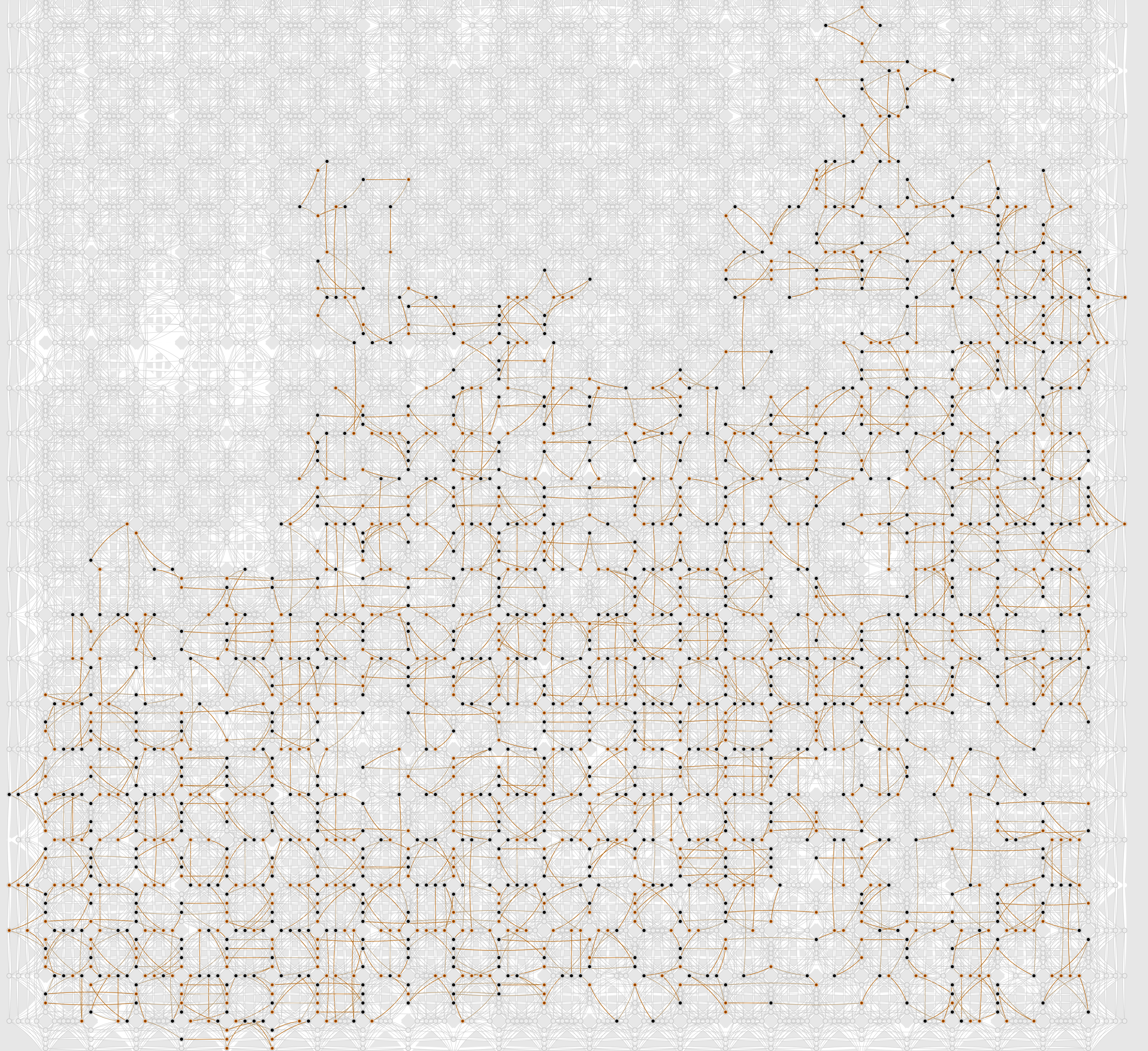}
\caption{Visualization of the Zephyr topology of the D-Wave Advantage2 quantum processing unit (QPU) with more than 4,400 qubits. Active qubits (orange and black dots) indicate spin orientations of +1 and -1 in a computed solution. The embedding simultaneously includes four 160-node and four 320-node rings mapped onto the QPU connectivity graph, with couplers represented as orange lines.}
\label{fig:embedding}
\end{figure}

Figure~\ref{fig:embedding} shows the embedding of a 160-qubit Ising chain on D-Wave's Zephyr graph. 

\end{document}